\documentclass[prb,twocolumn,showpacs,superscriptaddress,amsmath,amssymb]{revtex4}

\usepackage{graphicx}%
\usepackage{dcolumn}%

\begin{document}

\title{Superconducting gap in the presence of bilayer splitting in underdoped (Pb,Bi)$_2$Sr$_2$CaCu$_2$O$_{8+\delta}$.}

\author{S. V. Borisenko}
\affiliation{Institute for Solid State Research, IFW-Dresden, P.O.Box 270116, D-01171 Dresden, Germany}

\author{A. A. Kordyuk}
\affiliation{Institute for Solid State Research, IFW-Dresden, P.O.Box 270116, D-01171 Dresden, Germany}
\affiliation{Institute of Metal Physics of National Academy of Sciences of Ukraine, 03142 Kyiv, Ukraine}

\author{T. K. Kim}
\author{S. Legner}
\author{K.~A.~Nenkov}
\author{M. Knupfer}
\author{M. S. Golden}
\altaffiliation[New address: ]{Van der Waals-Zeeman Institute, University of Amsterdam, The Netherlands}
\author{J. Fink}
\affiliation{Institute for Solid State Research, IFW-Dresden, P.O.Box 270116, D-01171 Dresden, Germany}

\author{H. Berger}
\affiliation{Institut de Physique Appliqu\'ee, Ecole Politechnique F\'ederale de Lausanne, CH-1015 Lausanne, Switzerland}

\author{R. Follath}
\affiliation{BESSY GmbH, Albert-Einstein-Strasse 15, 12489 Berlin, Germany}

\date{\today}
\begin{abstract}
The clearly resolved bilayer splitting in ARPES spectra of the underdoped Pb-Bi2212 compound rises the question of how the bonding and antibonding sheets of the Fermi surface are gapped in the superconducting state. Here we compare the superconducting gaps for both split components and show that within the experimental uncertainties they are identical. By tuning the relative intensity of the bonding and antibonding bands using different excitation conditions we determine the precise {\bf k}-dependence of the leading edge gap. Significant deviations from the simple cos($k_{x}$)-cos($k_{y}$) gap function for the studied doping level are detected.

\end{abstract}

\maketitle
One of the crucial steps in the evolution of our understanding of superconductors has been the discovery of the energy gap between the ground state and the quasiparticle excitations of the system. Photoemission is one of those experimental techiques which provided direct evidence for the existence of such a gap in the high-temperature superconducting cuprates \cite{Imer,Manzke,Olson}. Subsequent ARPES investigations \cite{Dessau,Wells,Shen} clearly demonstrated the anisotropic character of the gap which now is considered as a hallmark of the high temperature superconductors. Detailed studies of the  symmetry properties of the gap suggested a purely \textit{d}-wave character \cite{Tsuei} in case of the optimally \cite{Ding} and overdoped \cite{Mesot} samples, as well as noticeable deviations from the simplest d-wave scenario upon underdoping  \cite{Mesot}. 

More recently, ARPES has made another step forward - technical improvements of the method allowed to resolve additional features in the momentum and energy distributions of the photoelectrons which were not resolved before. Among the new achievments is the recent observation of the bilayer splitting (BS) in over- and underdoped Bi$_2$Sr$_2$CaCu$_2$O$_{8+\delta}$ and (Pb,Bi)$_2$Sr$_2$CaCu$_2$O$_{8+\delta}$ bilayer compounds \cite{Feng1, Chuang1, Kord1}. The presence of the BS implies that some of the conclusions based on the earlier ARPES data should be revised. As a result of such a revisit it was already established that, for example, the famous broadening of the ARPES features in momentum \cite{Kord1} and energy \cite{Feng2} could be at least partially accounted for by the bilayer splitting. Very recent investigations, in which differentiation between bonding and antibonding components is essential, imply that the more active part of the Brillouin zone in a sense of a coupling to other degrees of freedom is the antinodal region of the Fermi surface rather than the ($\pi$,0)-point itself \cite{Kord2,Gromko}. At the same time detection of the BS not only provides answers but also results in further new questions. One of such natural and straightforward questions in the light of these findings is 'how does the superconducting gap behave on both Fermi surface sheets?' Could it be that the superconductivity in Bi2212 is a sheet-dependent one as, for instance, is the case in 2H-NbSe$_2$ \cite{Science} or MgB$_2$ \cite{Giubileo}? Does the strong correlation between the so-called range parameter (related to the hopping integral \textit{t'}) and T$_c$$_{max}$, found recently \cite{Andersen} for nearly all hole-doped one-layer and \textit{bonding} subband of multilayer HTSC cuprates, really mean that the bonding component plays a more important role in superconductivity than the other?

In this Communication we make a start at answering these questions by measuring the superconducting gap in modulation-free Pb-Bi2212 single crystals. Exploiting the relative variation of the photoionisation matrix elements for the bilayer split components, we are able to independently determine the leading edge gaps (LEG) corresponding to the different sheets of the Fermi surface in the superconducting state. Our results suggest that both gaps are comparable as follows from the analysis of the "map of gaps" around the ($\pi$,0)-point and other points in the Brillouin zone. Choosing the excitation conditions in a way which results in the suppression of the photocurrent corresponding to the antibonding component we determine the precise \textbf{k}-dependence of the LEG for its bonding counterpart for which emission then is favorable. The observed anisotropy of the leading edge gap for the studied doping level deviates from the conventional behavior described by the simple cos($k_{x}$)-cos($k_{y}$) function.

The ARPES experiments were carried out using angle-multiplexing electron energy analyzers (SCIENTA SES-200 and SES-100). Spectra were recorded either using $h\nu=$21.218 eV photons from a He source, as described elsewhere \cite{Kord1,BorisPRB}, or using radiation from the U125/1-PGM beamline \cite{Follath} at the BESSY synchrotron radiation facility. The total energy resolution ranged from 10 meV (FWHM) at $h\nu=$ 21 eV to 20 meV at $h\nu=$ 50 eV using the linearly polarized light as determined for each excitation energy from the Fermi edge of polycrystalline gold. The angular resolution was kept below 0.2$^\circ$ both along and perpendicular to the analyzer entrance slit. Data shown in Fig.4 were taken with 0.2$^\circ$ x 0.3$^\circ$ angular resolution. All data were collected on two similar underdoped, modulation-free single crystals of Pb-Bi2212 ($T_c$=77K). To achieve underdoping the as-grown samples were annealed in Ar atmosphere at 450$^\circ$C during 24 hours with subsequent fast cooling.

\begin{figure}
\includegraphics[width=8.47cm]{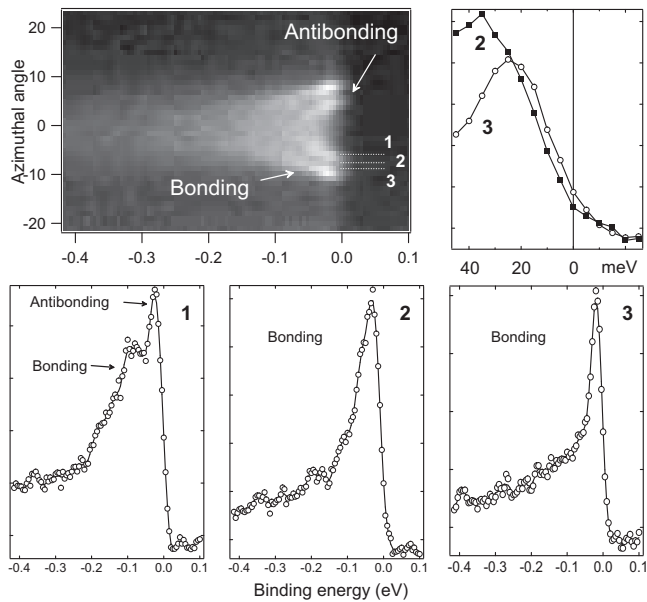}%
\caption{\label{Split} Energy distribution map recorded along the azimuthal path ($|{\bf k}|$=1.09 \AA$^{-1}$) at T=40K (upper left panel) showing the clearly resolved bilayer splitting. EDC's from the {\bf k}-points marked by the white dotted lines on the EDM (lower panels). Comparison of EDC's 2 and 3 on an expanded energy scale (upper right panel).}
\end{figure}

\begin{figure} 
\includegraphics[width=8.47cm]{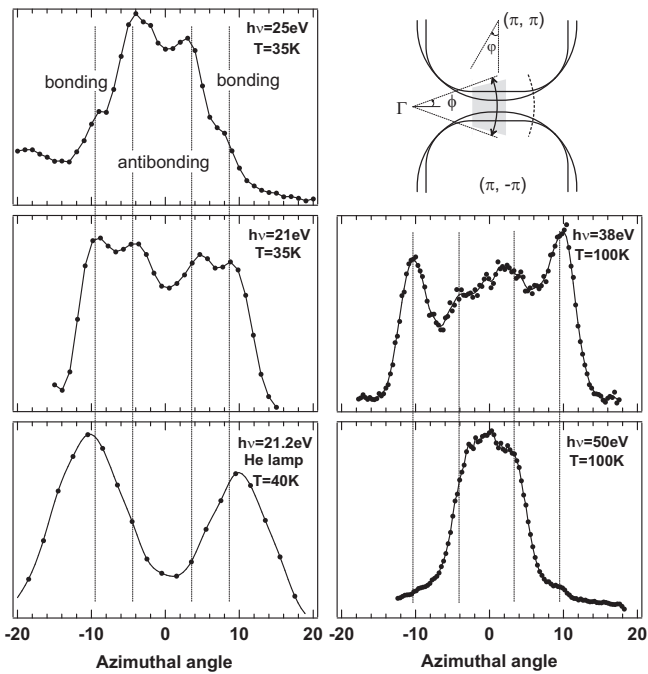}%
\caption{\label{MDCs} Azimuthal MDC's showing how relative intensity of the bonding and antibonding components could be tuned by choosing a suitable excitation energy. Inset illustrates which parts of {\bf k}-space were probed. Vertical dotted line show the angular positions of the bonding and antibonding antinodal points.}
\end{figure}

We start by presenting in Fig.~\ref{Split} what motivated us to initiate the present study. The explicitely seen bilayer splitting in the energy distribution map (EDM) recorded along the azimuthal path (see dashed line in the inset to Fig.~\ref{MDCs}) apparently signals the possibility of existence of the two gaps related to the bonding and antibonding Fermi surfaces. In the lower panels of Fig.~\ref{Split} we show three energy distribution curves (EDC's), which correspond to cuts located at the numbered white dotted lines on the EDM. Spectra 1 and 3 are taken at what would be {\bf k}$_F$ vectors if the system was ungapped and, despite being very close to one another in {\bf k}-space, they have completely different lineshapes from which a superconducting gap value can be derived. EDC number 2 is from just in between the gapped sheets of the FS and therefore is not suitable for the determination of the superconducting gap. The upper right panel of Fig.~\ref{Split} compares EDC's 2 and 3 in detail to illustrate how erroneous the determination of the superconducting gap could be if the precise {\bf k}-location is uncertain. In this case the peaks of the EDC's are separated by $\sim$10meV and the leading edges by $\sim$6meV. Depending on which method is used for the determination of the gap value (see below), the error could amount to as much as 50$\%$ of the gap value reported before for this {\bf k}-space region. It is also easy to predict an error which is the result of the non-resolved BS due to an insufficient angular resolution. Summing up EDC's from the larger {\bf k}-area would inevitably result in an overestimation of the gap values since both peak and leading edge positions of "irrelevant" EDC's are always at higher binding energies.

The necessity, then, to determine the superconducting gaps of both Fermi surface sheets is obvious. The only question remaining is 'how best to do this?' The problem here is that the situation depicted in Fig.~\ref{Split} is rather unique - it is in fact quite unusual to see bonding and antibonding bands equally well resolved. To achieve such a clarity, EDM shown in Fig.~\ref{Split} was recorded along the specific path in {\bf k}-space: both Fermi surfaces are cut at approximately right angles at points which are sufficiently away from the node (where BS vanishes) and antinodes where antibonding saddle point is close to the Fermi level and, due to the underdoping, superconducting gaps are large hampering the visual resolution of the BS. In most cases excitation conditions which define the distribution of the photocurrent via matrix elements are such that the bilayer splitting is hardly visible \cite{Chuang2,Kord1}. This is most likely the reason why the BS remained elusive for so long. Every change of a given excitation energy and geometry of the experiment leads to a particular, as a rule very strong, {\bf k}-dependence of the matrix elements which is, generally speaking, not the same for the bilayer split bands. It has been shown recently that the key factor defining relative bonding/antibonding intensity is the energy of the exciting photons \cite{Chuang2,Feng2,Kord2}. We exploit these observations here and search for suitable excitation conditions by varying the energy of the tunable synchrotron radiation. The most convenient region of the {\bf k}-space for such an exercise (given modulation-free crystals) is the vicinity of the ($\pi$,0)-point containing the antinodal points of the normal state Fermi surface: here both the gap and the bilayer splitting are expected to be largest.

In Fig.~\ref{MDCs} we present azimuthal momentum distribution curves (MDC) for the path indicated in the inset by the double-headed arrow. The MDC signal represents an integration over 1 degree of polar angle and the first 20 meV in binding energy. As expected, variations in the lineshapes are dramatic. If using linearly polarized radiation of h$\nu$=21eV or h$\nu$=38eV emissions from the bonding and antibonding bands are comparable, for h$\nu$=25eV and h$\nu$=50eV the dominance of antibonding component is obvious. At first sight, it may appear surprising that the MDC's taken with h$\nu$=21eV and h$\nu$=21.2eV are so different. The two important differences here are polarization and energy. Firstly, for the former, linearly polarized light was used (E$\parallel${\bf k}$_{\parallel}$), whereas the latter involved majority unpolarized radiation from a helium resonance source. Secondly, both theoretical and experimental studies have shown that there is a sharp local minimum in the matrix elements near $\sim$20 eV \cite{Lindroos,Kord2,Durr,Lee}, and thus in this region, even a small change in energy could be enough to provoke such a large variation of the photoemission signal.

At this stage we have to decide which energy would be the most suitable for the determination of the superconducting gaps related to both Fermi surfaces. This choice, in turn, depends on how we plan to extract the gap from the spectra. There are two widely-known approaches. The first involves fitting an EDC with a model spectral function multiplied by the Fermi function and convoluted with the energy and angular resolution functions \cite{DingPRL95}. The advantage of this method is that if the fit is successful, one gains access to the absolute value of a gap. The shortcomings of this method include increased error bars due to the fitting procedure itself and, more importantly, a lack of knowledge as regards the precise form of the spectral function, thus making this approach model dependent. The second possibility is the so-called leading edge gap method, in which the lowest binding energy where the {\bf k}$_F$-EDC looses half of its maximal intensity is determined. This approach is attractive in its precision (nowadays better than 1 meV) and simplicity and is by far the most widely applied in the HTSC ARPES literature. The drawback is that the LEG is only a good qualitative measure of the gap and does not yield its absolute value. Now both of these gap methods face a severe challenge when two separate, gapped features are in close proximity. In this context, the EDC fitting procedure would benefit from having both features of equal intensity (as, for instance, is the case when using linearly polarized 21eV photons, see Fig.~\ref{MDCs}), whereas the LEG determination would work better when the intensity from the antibonding component dominates, as then the weak bonding component will not influence the position of the antibonding leading edge unduly (see also Fig.~\ref{Split}). As the aim of this paper is a {\it comparative} study of the relative sizes and symmetries of the superconducting gaps of the two Fermi surface sheets, the LEG approach is the method of choice and thus the 25 eV data are the most appropriate to address this question.

Having chosen the excitation energy and the gap determination method we are left with the known problem of which particular EDC has to be used to define the LEG (see Fig.~\ref{Split}). Here we apply a mapping procedure allowing to cover a relatively large {\bf k}-space region and introduce thereby a new method of presenting information derived from ARPES spectra - "the map of gaps" - which is the momentum (or angular) distribution of the binding energies of the leading edges.

\begin{figure} 
\includegraphics[width=8.47cm]{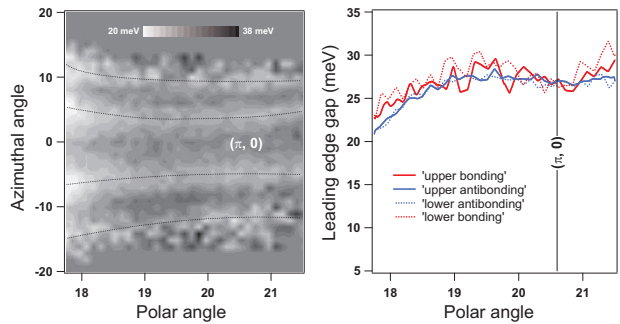}%
\caption{\label{Map25} Left panel: map of gaps derived from the EDC's taken with 25 eV photons at {\bf k}-points within the shaded area shown in the inset to Fig.\ref{MDCs}. Both of the gapped Fermi surface sheets are seen. Right panel: leading edge gaps as a function of the absolute {\bf k}-value. Zero gap corresponds to the binding energy (-7 meV) of the leading edge of the nodal {\bf k}$_F$- EDC.}
\end{figure}

In Fig.~\ref{Map25} we show such a map which is a grey scale image of leading edge binding energy. Four traces of the gapped Fermi surfaces are easily seen, despite the blurring of the bonding components as can be expected from the relative intensities seen in the 25 eV panel of Fig.~\ref{MDCs}. As a next step, we have determined the lines formed by joining the "minimum gap loci" \cite{DingPRL97}, which are then shown superimposed upon the map of gaps. The LEG values along these minimum gap loci are plotted in the right panel of Fig.~\ref{Map25}. Thus, as a first major result, these curves unambiguously demonstrate that the superconducting energy gaps on both Fermi surface sheets are identical, indicating no severe sheet dependence of superconductivity in this case.

Now, being certain that the gaps are identical, we can take a further step in trying to obtain information about their symmetry. To increase the accuracy of the measurements one can reduce now the influence of one of the bands assuming that both gaps have also the same {\bf k}-dependence in the other part of the BZ, i.e. monotonically decrease when approaching the nodal direction. Looking again at Fig.~\ref{MDCs} one sees that possible candidates are the photon energies of 21.2 and 50 eV at which emission from the antibonding or bonding bands, respectively, is significantly suppressed. As mentioned before, every experiment using linearly polarized radiation implies particular symmetry conditions which uniquely define the photocurrent distribution. In cuprates, due to the symmetry of the states forming the near-$E_F$ electronic structure, it is known that favorable conditions for the emission along the $\Gamma - (\pi, \pi)$ direction are such that the vector of polarization should be perpendicular to this direction whereas for $\Gamma - (\pi, 0)$ direction it has to be parallel  \cite{Dessau_Thesis,Durr}. In our experimental geometry, when using the linearly polarized excitation the polarization vector is always parallel to the studied direction in the reciprocal space and therefore cannot be equally effective for all states at the Fermi level. This implies that to be sure not to loose the spectral weight due to symmetry reasons, which is especially important when determining LEG, it is necessary to probe every {\bf k}-point with two mutually perpendicular orientations of the polarization vector which is technically quite difficult. For this reason we present here data taken with 21.2 eV radiation from the helium lamp which is mostly unpolarized. 

\begin{figure} 
\includegraphics[width=8.47cm]{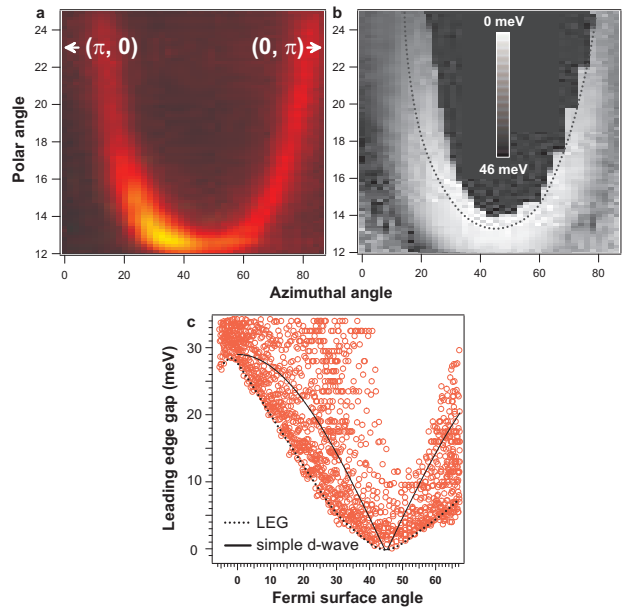}
\caption{\label{Map&LEG} a) Angular distribution map of the photoemission intensity integrated within 50 meV of $E_F$ showing the locus of {\bf k}$_F$-points. b) Map of gaps derived from the EDC's taken with 21.2 eV photons within the similar {\bf k}-space area. Both maps are presented in angular space and therefore give a slightly unfamiliar view of the Fermi surface. c) Plot of the leading edge gaps as a function of the angle around the Fermi surface. The red symbols are the data from panel (b), dotted lines in both (b) and (c) show the minimum gap locus and corresponding LEG values respectively, the solid line in (c) is the simple d-wave gap function $\Delta (\varphi)=\Delta_{max}|cos(2\varphi)|$.}
\end{figure}

Precise determination of {\bf k}-dependence of the superconducting gap and thus its symmetry requires precise navigation in {\bf k}-space in order to accurately locate a position on the (former) Fermi surface \cite{position}. For this purpose we show in Figure \ref{Map&LEG}a a map of intensity integrated within 50 meV energy interval which facilitates identification of the {\bf k}-vectors. We note here that not only this map but also an additional Fermi surface map taken at room temperature covering much larger area in {\bf k}-space were needed to assure accurate definition of the angle $\varphi$ \cite{position}.
Fig.~\ref{Map&LEG}b shows the greyscale "map of gaps". Visual inspection of the map of gaps already clearly points to the anisotropic character of the gap.

One can now use two approaches to construct the LEG($\varphi$) function. One method would be to define the path in the {\bf k}-space using, for example, maxima of MDC's  taken perpendicular to the FS and then plot the value of the LEG along this path as a function of $\varphi$. We show here the results of an alternative method which is based on the "minimum gap locus", but point out that both LEG($\varphi$) curves qualitatively agree. Adopting this second procedure, we simply replot  the same dataset (Fig.~\ref{Map&LEG}b) in other coordinates: LEG value versus Fermi surface angle ( Fig.~\ref{Map&LEG}c). This gives the red symbols shown in Fig.~\ref{Map&LEG}c. The curve joining the low-gap extremity of these data points represents the true {\bf k}-dependence of the superconducting LEG, and is shown as a dotted line in Fig.~\ref{Map&LEG}c. In this way, these extremal LEG values correspond to the points on the "minimum gap locus" dotted line in Fig.~\ref{Map&LEG}b).

As is clear from the comparison shown in Fig.~\ref{Map&LEG}c, that the obtained {\bf k}-dependence of the LEG for these UD77K crystals is quite
different from that expected in the case of a simple d-wave gap function cos($k_{x}$)-cos($k_{y}$). The most striking difference is that the gap behaves much flatter around the node resulting in a "U"-like shape rather than "V"-shape for LEG($\varphi$) function. Keeping in mind that LEG is only qualitative measure of a real gap it is crucial to understand to which extend LEG($\varphi$) reproduces the main features of a real gap function. According to our numerical simulations \cite{Kord_gap} measured binding energy of the nodal {\bf k}$_F$-EDC leading edge (-7 meV) corresponds to the \textit{gapless} spectral function thus implying the consistency of the presented data with the d-wave symmetry of the superconducting gap. It is also shown \cite{Kord_gap} that due to the temperature effects the LEG method tracks the real gap with an uncertainty of order of 2 meV (for the experimental conditions in question) when the corresponding absolute values of the real gap are of the same order. This last remark lives finite the probabilty that a nodal cusp can be present in the real gap function (although much more obtuse than is required by the modulus of the simplest d-wave function).

Observed partial suppression of the superconducting gap with respect to the simplest d-wave expectations in underdoped sample is in qualitative agreement with previous ARPES results \cite{Mesot} where authors extracted gap values by modeling the spectra taken at a few {\bf k}-points using a simple BCS spectral function without considering BS. Our high resolution data provide much more presice experimental information for a further quantitative analysis and support the results reported in Ref.~\onlinecite{Mesot} in a way that as it is clear now, using the 22 eV photons (as in Ref.~\onlinecite{Mesot}) one is sensitive mostly to the bonding component \cite{Kord2} and determination of the superconducting gap is not influenced by the presence of the BS. Our simulations \cite{Kord_gap} show that the relation between the LEG and the real gap function is nontrivial and model-dependent and therefore we do not make an attempt here to extract quantitative information following authors of Ref.~\onlinecite{Mesot} but share their opinion that the observed behavior could be explained invoking higher harmonics \cite{harmonics} of the gap and is probably connected with increased range of the pairing interaction in underdoped compounds due to reduced screening (see also \cite{Chernyshev, Ghosh}). In this context we mention related theoretical work \cite{Haas} where momentum-dependent scattering from the impurities may also lead to extended gapless regions in the gap function centered around the nodes of the pure d-wave superconductor. 

As another possible scenario we mention finally that a preliminary analysis of ARPES spectra from the same samples taken above T$_c$ suggests that the {\bf k}-dependence of the pseudogap is similar to that of the superconducting gap presented here. However, at this stage, it is too early to draw conclusions from this as to what extent the normal state and superconducting state gaps may share the same microscopic origin.

In conclusion, we have demonstrated that in underdoped, modulation-free Pb-Bi2212 each of the bilayer split Fermi surface sheets supports a gap in the superconducting state. We compared the gap values in the antinodal region where both are expected to be maximal and found that the gap values on
both sheets are identical within the experimental error bars. By varying the experimental excitation conditions we could tune the relative intensity from the bonding/antibonding states and thus, via suppression of the emission from the antibonding band, we were able to precisely define the symmetry of the leading edge superconducting gap for the bonding Fermi surface sheet. Significant deviations from the simplest cos($k_{x}$)-cos($k_{y}$) d-wave behavior were observed, with a strong flattening out of the gap function around the Brillouin zone diagonal.

We acknowledge stimulating discussions with H. Eschrig, S. Shulga, S.-L. Drechsler and M. Eschrig and technical support from R. H\"ubel. We are grateful to the DFG (Graduiertenkolleg "Struktur- und Korrelationseffekte in Festk\"orpern" der TU-Dresden) and to the Fonds National Suisse de la Recherche Scientifique for support.


\begin{thebibliography}{}
\bibitem{position} The common framework in which to discuss the gap symmtery is as a function of the angle $\varphi$ between the ($\pi, \pi$)-($\pi$, 0) line and the ($\pi, \pi$)-FS directions in the reciprocal space, where FS represents the normal state Fermi surface. $\varphi$ is connected with the standard ARPES azimuthal ($\phi$) and polar ($\vartheta$) angles via the simple relation: ${\cos}^2\varphi={(1+{((1-\eta\sin\vartheta\cos\phi)/(1-\eta\sin\vartheta\sin\phi))}^2})^{-1}$, where $\eta$ is defined by the kinetic energy of the photoelectrons and the lattice constant ($\eta\sim 2.58$ when using 21.2 eV photons).
\end{thebibliography}
\end{document}